\documentclass{comnet}
\usepackage{url}
\usepackage{xcolor}
\usepackage{diagbox}
\usepackage{multirow}
\usepackage{makecell}
\usepackage{blindtext}
\usepackage{graphicx}

\begin{document}

\title{BFS based distributed algorithm for parallel local directed sub-graph enumeration}

\shorttitle{Efficient GPU based local motif counting}

\author{
\name{Itay Levinas}
\address{Bar Ilan University, Ramat Gan, Israel }
\name{Roy Scherz}
\address{Bar Ilan University, Ramat Gan, Israel}
\and
\name{Yoram Louzoun$^*$}
\address{Bar Ilan University, Ramat Gan, Israel\email{$^*$Corresponding author: louzouy@math.biu.ac.il}}}

\maketitle

\begin{abstract}
{Estimating the frequency of sub-graphs is of importance for many tasks, including sub-graph isomorphism, kernel-based anomaly detection, and network structure analysis. While multiple algorithms were proposed for full enumeration or sampling-based estimates, these methods fail in very large graphs.  Recent advances in parallelization allow for estimates of total sub-graphs counts in very large graphs. The task of counting the frequency of each sub-graph associated with each vertex also received excellent solutions for undirected graphs. However, there is currently no good solution for very large directed graphs.

We here propose VDMC (Vertex specific Distributed Motif Counting) - a fully distributed algorithm to optimally count all the 3 and 4 vertices connected directed graphs (sub-graph motifs) associated with each vertex of a graph. VDMC counts each motif only once and its efficacy is linear in the number of counted motifs. It is fully parallelized to be efficient in GPU-based computation. 

VDMC is based on three main elements: 1) Ordering the vertices and only counting motifs containing increasing order vertices, 2) sub-ordering motifs based on the average length of the BFS composing the motif, and 3) removing isomorphisms only once for the entire graph. 
We here compare VDMC to analytical estimates of the expected number of motifs and show its accuracy. VDMC is available as a highly efficient CPU and GPU code with a novel data structure for efficient graph manipulation. We show the efficacy of VDMC and real-world graphs.

VDMC allows for the precise analysis of sub-graph frequency around each vertex in large graphs and opens the way for the extension of methods until now limited to graphs of thousands of edges to graphs with millions of edges and above. 

GIT: https://github.com/louzounlab/graph-measures}
{Subgraphs, BFS, GPU, Full enumeration}
\end{abstract}

\section{Introduction}
A common and important task in graph (network) analysis is the frequency of sub-graph analysis. This task has two versions: counting non-overlapping subgraphs (\cite{elhesha2016identification}), and counting all subgraphs (\cite{ribeiro2019survey}). 
Such counting is important for multiple theoretical and computational tasks, including among many others: Sub-graph isomorphism (\cite{ullmann1976algorithm}), graph classification (\cite{jin2010gaia}), graph anomaly detection (\cite{papadimitriou2010web}), and clique counting (\cite{bron1973algorithm}).  Within sub-graph analysis, a special focus has been given to the frequency of small connected sub-graphs, with typically 3-4 vertices. Such sub-graphs are often called graphlets (\cite{ahmed2015efficient}), or sub-graph motifs (\cite{kashtan2004efficient}). In a directed graph, motifs are required to be connected only in the underlying undirected graph (e.g. $A\rightarrow B$, $A\rightarrow C$ is a motif, although there is no path from $B$ to $C$). A large number of algorithms was proposed to count sub-graphs. Some algorithms are specific to a certain type of sub-graphs (e.g. cliques (\cite{bron1973algorithm}), triads (\cite{schank2005finding}) or stars (\cite{gonen2011counting})), but most existing algorithms focus on counting motifs, either directed or undirected (For an excellent recent review see  \cite{ribeiro2019survey}).  
Existing methods to count motifs can be divided into four main groups: Full counting methods (e.g. among many others  \cite{itzhack2007optimal,wernicke2006fanmod}) and sampling methods (See  \cite{wang2017moss,yang2018ssrw,bressan2018motif} for a few recent examples among many others) for directed and undirected motifs. Beyond those, there are currently over 50 existing algorithms for motif counting. Both sampling and full enumeration methods provide accurate estimates of the total sub-graph number for any given motif.

A related work also extensively studied especially in undirected graphs is the count of the sub-graphs frequency around each vertex. This frequency is useful for many applications, including topology-based machine learning algorithms (\cite{naaman2019edge}).

Formally, local subgraph counting is defined as the number of times a given graph $p$ exists around a given vertex. 
Such local subgraph counting (or local motifs in the case of graphs connected in the underlying undirected graph) has various applications, including among many others: vertex clustering (\cite{yin2018higher}), the detection and characterization of dense communities (\cite{tsourakakis2015k}), network characterization and classification (\cite{prvzulj2007biological,benami2019topological,naaman2019edge}), and also applications to machine learning, where these motifs are used as the input to classification and anomaly detection tasks (\cite{hayes2013graphlet,akoglu2015graph}). Such local motif counts have also been for link prediction (\cite{rossi2012transforming,naaman2019edge}). 

The local sub-graphs studied vary among applications. However, many of them focus on undirected connected sub-graphs with a pre-defined number of vertices (further denoted here as undirected $k-motifs$, where $k$ is the number of vertices). 
Recently, local subgraph counting algorithms have been extended to large graphs (\cite{ahmed2016estimation,elenberg2015beyond,hovcevar2014combinatorial,melckenbeeck2016algorithm,ahmed2016exact,elenberg2016distributed,pashanasangi2020efficiently,melckenbeeck2018efficiently}\cite{pashanasangi2020efficiently}). Note that the position of the vertex in the sub-graph is usually not taken into consideration. One can define three main approaches to local motif counts in large graphs: decomposition,  matrix-based methods and enumeration methods. The first two methods are limited to undirected sub-graphs (even if the original graph is directed, one can count undirected sub-graph in the undirected graph induced by ignoring the direction of edges in the original graph), while the last can be applied to both directed and undirected sub-graphs. Specifically, for a given subgraph type $p$ and a graph $G$:

\begin{itemize}
\item In matrix based approaches (\cite{hovcevar2014combinatorial,hovcevar2017combinatorial,melckenbeeck2018efficiently,dave2017clog}), one counts sub-graphs by solving linear equations relating the numeration of larger motifs (higher $k$) to combinations of smaller motifs. 
\item Decomposition-based approach (\cite{elenberg2016distributed,pashanasangi2020efficiently}) are similar to matrix based approach, but simpler, since they only correlate $k-motifs$ to smaller components of the same motifs. 
\item Enumeration approaches (\cite{park2016pte,wang2019benu}) are counts by enumerating all matches of $p$ in $G$.  This is obviously the most computationally expensive method, but it is not limited to specific types of graphs in undirected graphs. 
\end{itemize}

Still, there is currently no efficient enough fully parallel enumeration approach to count directed local-motifs. Moreover, most methods focus on the detection of a single motif, instead of counting all possible motifs simultaneously.

We here provide a solution for efficient and fully parallelized counting of all local sub-graphs, and propose an algorithm to count the number of each motif that contains each vertex or edge, named VDMC (Vertex specific Distributed Motif Counting). In the following sections, we first outline the algorithm and then prove the main claims underlying the algorithm. We then provide an analytic estimate of the frequency of motifs in Erd\H{o}s R\'enyi graphs (\cite{erdHos1960evolution}) and show that VDMC produces the expected motif frequency. We discuss the VDMC performance on large random and real world graphs in both CPU and GPU machines. 
\section{Novelty}
VMDC uses multiple elements, most of which have already been used in different algorithms, but never combined to produce a  highly efficient parallel vertex-specific sub-graph counting algorithm: 
\begin{itemize}
\item \textbf{Explicit counting.} Explicitly counting the number of each sub-graph type containing a given vertex, through a BFS on the undirected graph underlying the directed graph (i.e. the graph produced by ignoring the direction of edges). 
\item \textbf{Divide and conquer.} Applying a divide and conquer approach, by (de-facto) removing from the graph each vertex for which we computed all motifs containing it. Each vertex is assigned an index based on its degree. Vertices are ordered, from the highest degree to the lowest, with an arbitrary order between vertices of equal degree. In each sub-graph, only motifs with indices higher than the root are counted. The computation of sub-graphs is performed in parallel, based on the lowest index in the sub-graph. This order makes the parallelization more efficient since low index roots contain more motifs than high index roots.
These low index roots are removed from the graph at the beginning which prevents re-passing on these "heavy" roots.Such a balancing optimizes the parallelization efficacy.
\item \textbf{Efficient data structure.} VDMC uses the cache-aware efficient Compressed Sparse Row(CSR) format (\cite{bulucc2009parallel}) that minimizes the memory cost to the number of edges and allows for cache-aware parallelization.
\item \textbf{A set of efficient rules to count each sub-graph once and only once}, as will be further explained in lemmas bellow.
\item \textbf{Ordering possible motifs based on the average distance in the BFS of all vertices}. To count all motifs only once, VDMC counts all motifs that have a structure with a low average BFS depth before motifs in structures with a higher BFS depth.
\item \textbf{Combining isomorphisms only once at the end of the counting process.} During the counting equivalent motifs are counted separately. Only at the end of the enumeration, isomorphic motifs are combined.
\item \textbf{GPU implementation to maximize parallelization}.
\end{itemize}
\section{Related Work}
Beyond the vast number of motif counting algorithm, multiple parallel enumeration algorithms were developed. Those are typically based on either Map-reduce applications, GPU, Shared, or Distributed Memory (SM or DM). Those include (a non-comprehensive list, mainly focused on the GPU based applications): (Wang et al  \cite{wang2005parallel}, Schatz et al  \cite{schatz2008parallel}, MPRF  \cite{liu2009mapreduce},Ribeiro and collaborators (multiple algorithms)  \cite{ribeiro2010parallel,ribeiro2010efficient,eddin2017scalable}, Elenberg et al \cite{elenberg2016distributed}, Rossi et al \cite{rossi2016leveraging}, Ahmed et al \cite{ahmed2015efficient}, Lin et al \cite{lin2016network} and Milinkovich at al \cite{milinkovic14contribution}). 

We follow \cite{wang2005parallel} in distributing vertices to get an equal work share, but instead of using a combination of high and low degree vertices for each processor to better balance the work, we here sort the vertices by their (undirected) degree. In the analysis of each vertex, we compute only the following index vertices. Thus, as the vertex index increase, so does their degree, but in parallel, they do not process vertices of lower index numbers. Our approach is most similar to the work of Lin et al (\cite{lin2016network}). The main extensions here are counting the number of motifs that contain each vertex, to allow for vertex-specific motif counts. Moreover, VDMC uses a highly efficient memory usage formalism to reduce the total memory cost and the access time to memory in both CPU and GPU based applications. 

\section{Notation and Detailed description of algorithm}
\subsection{Definitions}
We propose an algorithm that counts all motifs once and only once in parallel computation. We first suggest the algorithm notation and describe the algorithm. We then prove that this algorithm counts each motif once and only once. We use the following notations for an unweighted directed graph $G=(V, E)$, where $V$ are vertices and $E$ are edges, and arbitrary indexing of the vertices, where vertex $i$ is the vertex with index $i$. Without loss of generality,  we assume that $i$ ranges between 1 and $|V|$.
\begin{itemize}
\item We denote by $G_U$ the undirected graph underlying $G$ produced by ignoring the direction of the edge in $E$. 
\item A sub-graph is a $k-motif$ if it is composed of $k$ vertices, and these vertices are a connected sub-graph of $G_U$. 
\item Each $k-motif$ is fully determined by the vertices composing it and can be described as a BFS based tree in $G_U$ (since it is connected), starting from a root $i$, which is one of the $k$ vertices. The BFS tree is denoted as a $k-BFS$. We denote the depth of an edge as its minimal distance from the root $d_i$, and the average depth of the motif as the average depth of the vertices composing it. For example, a $4-motif$, with the root connected to three vertices has an average depth of $0.75$, since the root has a depth of 0, and all other vertices have a depth of 1.
\item A $k-BFS$ will be defined to be proper if the index of the root is smaller than all other indices in the $k-BFS$.
\item We denote all $k-BFS$ with a root index of $i$ as $k-BFS(i)$.
\item To classify a $k-BFS$ into a specific $k-motif$, we follow \cite{itzhack2007optimal}, and use two sets of $k-motif$ values. The first value is different for isomorphs of the same motif, and the second is the lowest value for all isomorphs of the same $k-motif$ (See Figure 1 for example). The index of a $k-motif$ is based on a bit array of adjacencies. The graph is induced by $k$ vertices in any given order, and the edges between them can be represented as a miniature adjacency matrix, with 1 in the $(i,j)$ position if an edge exists between $i$ and $j$. This adjacency matrix is then represented as a binary vector by ordering it by rows, and removing the diagonal (since we assume a simple graph with no self edges). This binary vector is then translated to a base 10 number, which is the index of the motif.

\end{itemize}
\begin{figure}
\centering\includegraphics[scale=0.2]{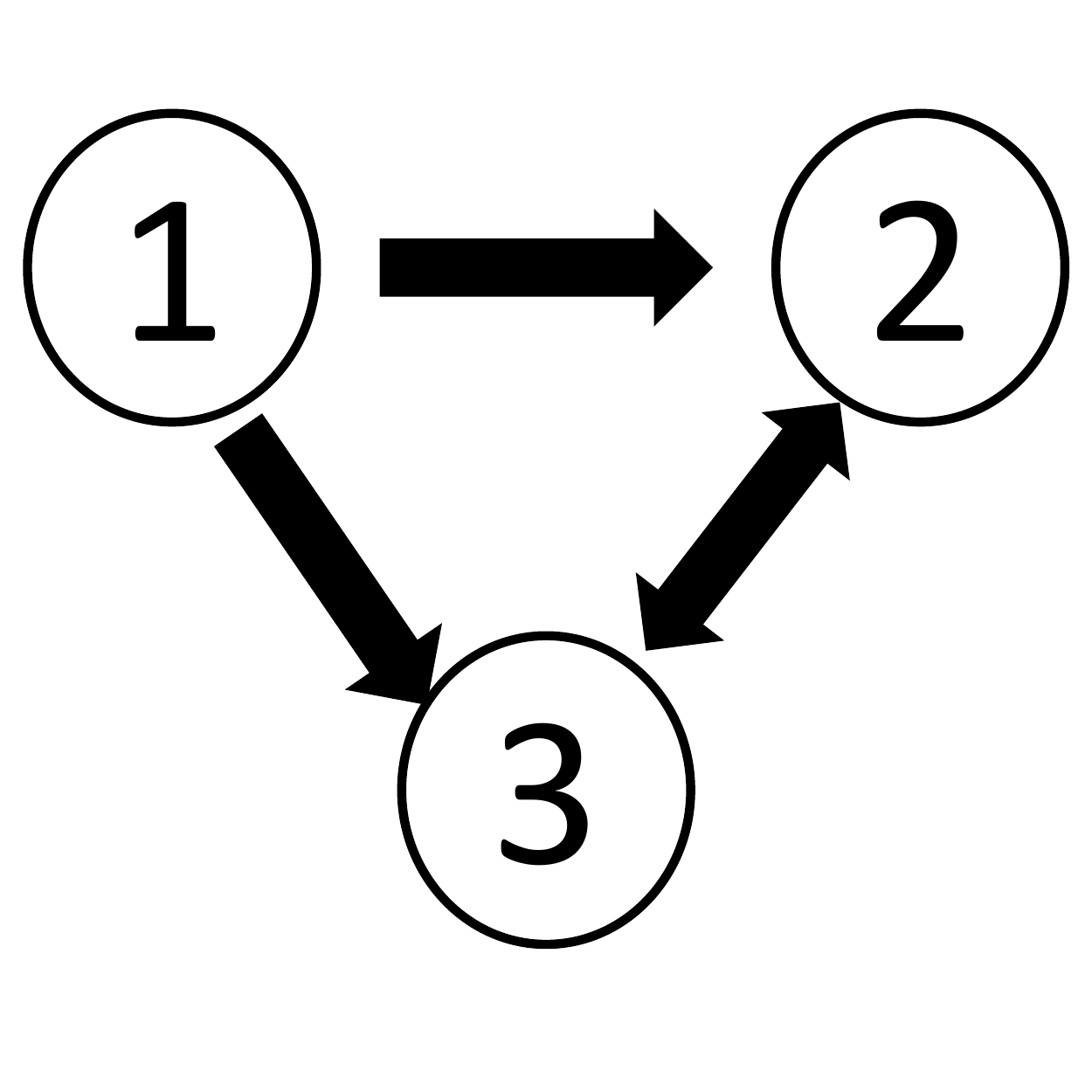}

$
matrix=\left(\begin{array}{ccc}
- & 1 & 1\\
0 & - & 1\\
0 & 1 & -
\end{array}\right) \Rightarrow bitstring=110101 \Rightarrow index=53 \Rightarrow index_{Min}=30
$

\caption{Motif indexing example. A $3-motif$ with 4 edges. The motif in the plot is translated to an adjacency matrix, which is in turn translated to a bit string. The bit string is translated to a base 10 number, which is the index of the motif. At the end of the analysis all isomorphic motifs are summed, and their index is the minimal index of all isomorphs.}
\end{figure}

\subsection{The Algorithm}

\begin{figure}
\includegraphics[width=0.9\textwidth]{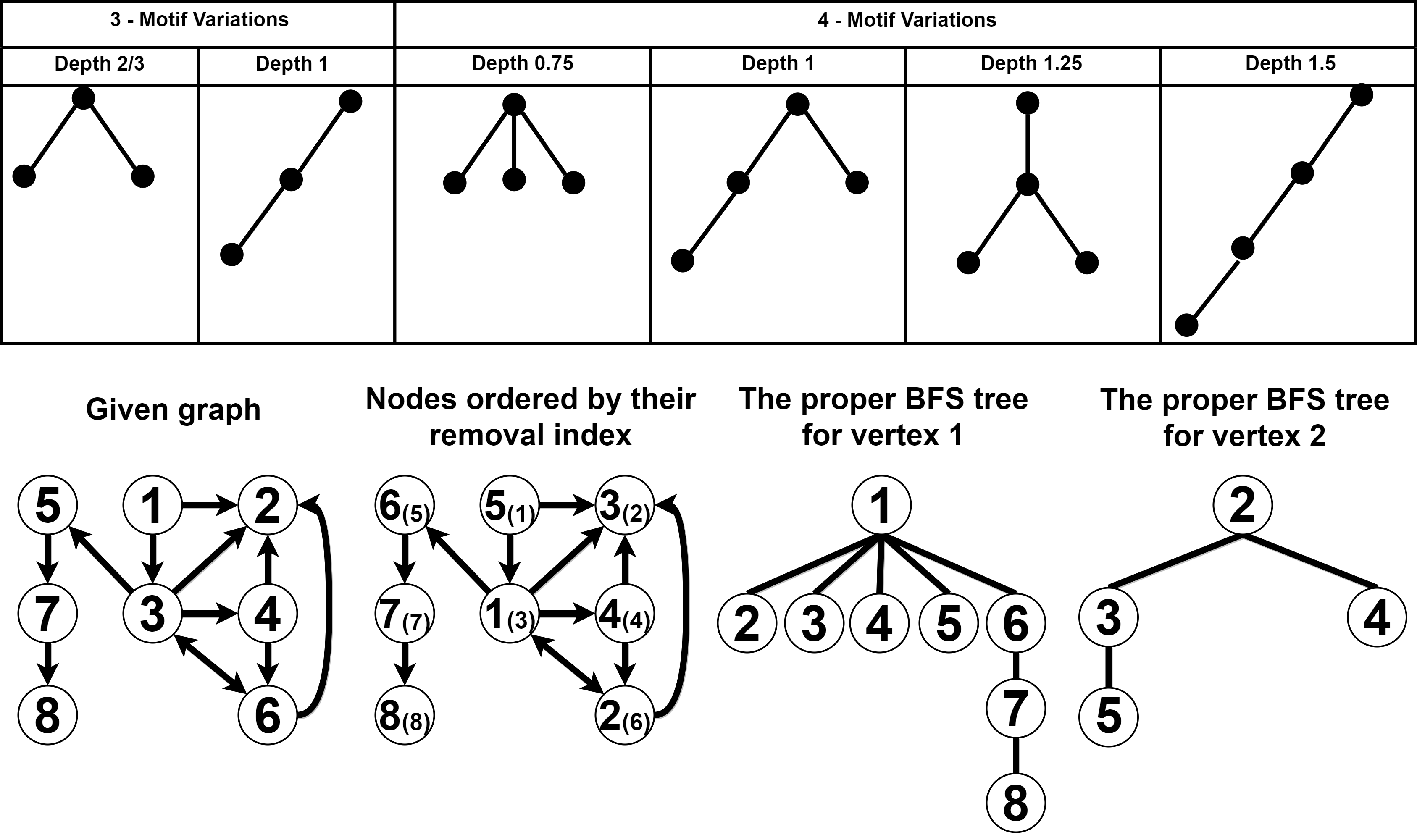}
\caption{First row: 3- and $4-motif$ variations - From left to right: $3-motif$ of depths $2/3$ and $1$, 4-motifs of depth $0.75,1,1.25$ and $1.5$. Second row: Visualization of The algorithm for a given graph. From left to right: the graph itself, the ordered vertices, the proper BFS tree for vertex number 1, the proper BFS tree for vertex number 2.}
\label{fig:example}
\end{figure}

Using these notations, and the proof below (in Section \ref{proofs}), we propose the VDMC algorithm:
\begin{itemize}
\item Sort all vertices in any arbitrary order. 

\item Set the number of motifs of each type (including all different isomorphs) for each vertex to be 0.
\item Split all $k-BFS(i)$ to parallelize the analysis.
 \begin{itemize}
   \item For each vertex $i$ compute all proper $k-BFS(i)$ and update the motif count for each vertex included in each $k-BFS(i)$, as follows.
 \begin{itemize}

\item For each such $k-BFS$ compute the depth of each vertex in the BFS, order vertices by their depth, and then by their index. (See Figure \ref{fig:example} for example).
\item Compute the adjacency matrix of this $k-BFS$ and compute the motif type it creates. For example in Figure \ref{fig:example}: 1-2-3-4, 1-2-6-7, 1-6-7-8 are 4-motifs of depth 0.75, 1 and 1.5 respectively.
\item Increase the counter for each vertex in the $k-BFS$ for this motif type by 1.
\end{itemize}
\end{itemize}
\item Sum over all isomorphs of each motif (See further description below).
\end{itemize}

To further improve the algorithm efficacy, we use in the  C++ kernel the highly efficient CSR format for sparse graphs. This format increases the code efficacy by leveraging the computer's internal cache mechanism. The CSR is composed of two arrays:
\begin{enumerate}
\item A consecutive ordered list of all the ordered lists of each vertices neighbors (Neighbors), maintained as a single array.
\item The starting index of each vertex's neighbor list in the Neighbor array (Indices).
\end{enumerate}
To understand the CSR format, here is a simple example. Assume the following graph: ($0\rightarrow1, 0\rightarrow2, 0\rightarrow3, 2\rightarrow0, 3\rightarrow1, 3\rightarrow2$).
\begin{itemize}
\item If the graph is directed, $Indices$ is: {[}0, 3, 3, 4, 6{]}, and $Neighbors$ is: {[}1, 2, 3, 0, 1, 2{]}.
\item If the graph is undirected, $Indices$ is: {[}0, 3, 5, 7, 10{]}, and $Neighbors$ is: {[}1, 2, 3, 0, 3, 0, 3, 0, 1, 2{]}.
\end{itemize}

The CSR object is designed around the principle of cache-awareness. When accessing the graph in the C++ code, the aim is to accelerate the computations by loading sections of the graph into the cache ahead of time for quick access. This comes into effect in the BFS, when we access the blocks of neighbor vertices in the Neighbors vector, and pulling the entire list of neighbors of a certain vertex into the cache.

During the motif enumeration a separate count is used for different isomorphs of the same motif. For example, the motif in figure 1 can be counted as 53 (110101) or as 30 (011110). Eventually, all isomorphs of the same motif are summed to the ones with the lowest index (either in real-time, or at the end of the analysis).

\section{Proof that each motif can be counted once and only once in parallel}
\label{proofs}
{\it \bf Lemma 1} If all and only proper $k-BFS(i)$ are counted, and within each $k-BFS(i)$ each $k-motif$ is counted once and only once, then each $k-motif$ is  counted once and only once. 

{\it \bf Proof} $k-motifs$ are fully determined by the vertices composing them. For any set of $k$ vertices ($X$), we can denote the minimal index as $j$. Since these vertices are connected in $G_U$, such a $k-motif$ will be counted within each $k-BFS(i) | i \in X$. However, only $k-BFS(j)$ is proper, and thus the motif composed of $X$ will be counted in $k-BFS(j)$ and only there. Since it will be counted there once and only once, each $k-motif$ will be counted once and only once.

{\it \bf Conclusion}. Counting all motifs can be performed by counting only all proper $k-BFS$, and can be parallelized by counting in parallel all motifs in all proper $k-BFS(i) | i \in V$. 

{\it \bf Lemma 2} In $3-motifs$, for a given $i$, there are only 2 possible structures for the $k-BFS$ and for the $4-motif$ there are only 4 possible $k-BFS$, as detailed in figure \ref{fig:example}. 

{\it \bf Proof}. This can be seen by enumerating all possibilities at the 1st level (2 for $3-motifs$, and 3 for $4-motifs$), and then the resulting possible combinations in lower levels.

{\it \bf Remark}. Each structure has a different average depth and can be defined by its depth.

Before introducing the next lemma, let us follow a given $k-BFS$. Such a $k-BFS$ has an average depth. A vertex in a $3-motif$ can only have vertices of depth 1 or 2 (beyond the root). A vertex in a $4-motif$ can be of depth 1,2 (with two variations) or 3 (Figure \ref{fig:example}). In the example in Figure \ref{fig:example}, the set of vertices ${1,3,4,5}$ can be counted six times in a $4-BFS(1)$ of depth 0.75 ($d=0.75, 1 \rightarrow 3,1 \rightarrow 4, 1 \rightarrow 5$,  $d=0.75, 1 \rightarrow 3,1 \rightarrow 5, 1 \rightarrow 4$, $d=0.75, 1 \rightarrow 4,1 \rightarrow 3, 1 \rightarrow 5$, etc.)
twice in a $4-BFS(1)$ of depth 1 ($d=1, 1 \rightarrow 3,1 \rightarrow 5, 3 \rightarrow 4$ and $d=1, 1 \rightarrow 5,1 \rightarrow 3, 3 \rightarrow 4$), and twice in a $4-BFS(1)$ of depth 1.5 ($d=1.5, 1 \rightarrow 4,4 \rightarrow 3, 3 \rightarrow 5$ and $d=1.5, 1 \rightarrow 5,5 \rightarrow 3, 3 \rightarrow 4$). To prevent, the double counting within the same level, one can simply determine that within the same level, the BFS should always follow the index number (i.e. not count $d=0.75, 1 \rightarrow 4,1 \rightarrow 3, 1 \rightarrow 5$). To prevent all counts of the $d=1, 1.5$, one can simply determine that no edges in the BFS are allowed from a given depth to a lower or equal depth.  Thus, if we do not count motifs that point within the same level from a high to a low count, and from a higher level to a lower or equal level, we will ensure that ${1,3,5,7}$ will be counted only once. This can be stated through the following lemma.

{\it \bf Lemma 3} If one does not count $k-motifs$ in a given structure with a direction from a higher depth to a lower or equal depth in the BFS, and among vertices of the same depth, vertices are only considered in order of their index, then no $k-motif$ will be counted twice in the same $k-BFS$.

{\it \bf Proof} Given a root $i$ and $k-1$ other vertices $X_1$. Each vertex is assigned a single depth, which is its minimal depth (i.e. if a vertex is both a first and second neighbor of the root, it is assigned the minimal depth, which is $1$). The two rules above induce a full order between all vertices in $X_1$ (depth and index). Each motif in $X_1$ is counted only following the order defined by the rules above, as such it will be counted only once. 

{\it \bf Note} Lemma 3 is meant to eliminate the case of counting twice in the motifs. This does not ensure that each motif is counted at least once, as will be further shown. However, we can explicitly state the conditions for $4-motif$ not to be counted and address that.

{\it \bf Lemma 4} Any  $4-motif$ not counted by the rules above is of average depth $1.5$ (4 vertices in a row) and is part of a loop containing the uncounted motif and a single extra external vertex.

{\it \bf Proof} Each $4-motif$ is per definition contained in a proper $4-BFS$. It will not be counted only if it is ignored following the conditions in Lemma 3. It cannot be ignored following two vertices of the same depth but in the wrong order, since the $4-BFS$ with the opposite order would be counted. Thus, the only condition that a $4-motif$ is not counted would be an edge in the BFS from a higher to a lower or equal depth vertex. Let us denote these vertices $v$ (lower depth - closer to root)  and $w$ (higher depth farther away from the root). $v$ cannot be of depth 0, per definition (since otherwise, the root would appear twice in the $4-BFS$). Thus, $v$ is at least of depth $1$. Similarly, $w$ cannot be of depth more than 2, since any vertex at depth 3 is the last vertex in a $4-BFS$ and does not point to anything. $v$ and $w$ cannot be of depth 1 in the same $4-BFS$ as explained by the symmetry argument above. Thus, $w$ must be of depth 2 and $v$ must be of depth 1 or 2. if $v$ would have been of depth 1, the proposed $4-BFS$  be counted in the average depth 1 $4-BFS$, leaving the last option of both $v$ and $w$ being of depth 2. They cannot be from the same $4-BFS$ (since they would be either with a common depth 1 ancestor, and then the symmetry argument above would hold, or from different ancestors, but then one would require at least 5 vertices in the $4-BFS$. One is only left with the choice that $v$ is of depth 2 but through a vertex not in the current $4-BFS$.

{\bf Comment} It is beyond the scope of the current manuscript, but a $k-BFS$ not accounted will contain a circle with a vertex outside this specific BFS but with the same root. The length of such a loop in the case of a $4-motif$ is exactly 5. Therefore we know that the problematic vertex was marked at both depths 2 and 3 exactly.

The counting algorithm takes this case into account and counts a $d=1.5$ motif even if the last node is marked of depth 2, as long as it is not from the same $k-BFS$. This actually simplifies the algorithm, since, for the last vertex, one must not check if it has any neighbor of depth 1, but rather if he is directly connected to the edge at depth $1$ in the current BFS.

\section{Vertex ordering and splitting over the GPU}

As mentioned above (in the definition of a proper k-BFS), each vertex is associated with a removal order. To achieve better performance, vertices are ordered in reverse order of the vertices' degrees, i.e. the vertex with the highest degree was associated with the lowest index, and the vertex with the lowest degree was associated with the highest index. This way, the motifs for the vertices with the highest degrees are calculated at the beginning and these vertices are removed from the graph. Hence there is no re-passing on these "heavy" vertices.

Real-world networks often have scale-free degree distribution, and as such may be computationally expensive. VDMC can handle very high degrees through the division of the $k-BFS$ for high degree vertices into parallel computations.
To parallelize the analysis, VDMC is adapted to GPU to analyze large blocks of vertices in parallel.
Each pair of a vertex and one of its neighbors is computed separately. Dividing the vertices on the GPU even by their neighbors was done to make the parallelization more efficient by making the blocks' tasks more equal. It prevents a situation where the algorithm waits only for a small number of vertices with a very high degree while the calculations for the other vertices have been completed. 

The following description is slightly technical, and can be skipped:

Two-dimensional blocks were created and arranged in a grid. Each block contained 16 threads for each dimension. The grid was also defined to be two-dimensional in order to calculate each pair of vertex and its neighbor in a separate block. The dimensions of the grid are $[(numOfNodes + blockSize.x - 1) / (blockSize.x),(numOfNodes + blockSize.y - 1) / (blockSize.y)]$ and each of them was limited to $2^{15}$ blocks (This number may  be modified according to the datasets and the memory available on the GPU). For datasets in which the number of blocks was larger than the defined limit, the excess vertices were divided again starting from the first block. That is, the BFS for the i-th vertex and its j-th neighbor was calculated in the block whose indices are $[i\%(grid\ size\ in\ x\ dimension),j\%(grid\ size\ in\ y\ dimension)]$.This ensures the blocks' tasks to be  even and a maximal parallelization of the analysis over the GPU threads.

\section{Comparison to Theory in $G(n,p)$}
To test the accuracy of VDMC, we compared the observed and expected number of each motif in random graphs $G(n, p)$ (also commonly named Erd\H{o}s R\'enyi graphs). We define a $G(n, p)$ as a graph with $n$ vertices, where each pair of vertices is connected with a constant probability $p$, independent on all edges or vertices. Let $X_{k, m}(i)$ be the number of $k-motifs$ of index $m$ that contain vertex $i\in V$.  We compute the expected value of $X_{k, m}(i)$ within an arbitrary $G(n, p)$ ( $\mathbb{E}\left[X_{k, m}(i)\right]$).

Denote $\mathcal{N}_{k}(i)$ as the set of all combinations of $k$ vertices that contain vertex $i$. For a specific combination of vertices $c \in \mathcal{N}_{k}(i)$, we denote the event "The induced sub-graph from $c$ is a motif of type $m$" as $A(c,m)$, the number of isomorphs of a motif of type $m$ as $N_{Iso}(m)$, and the number of edges in a motif of type $m$  as $n_{e}(m)$. The maximal number of edges within $k$ vertices is $n_{max}(k) = \begin{pmatrix} 
k \\ 2
\end{pmatrix}$ if the graph is undirected and twice that number otherwise. The indicator function is denoted as $\mathbb{I}\left[x\right]$

 $X_{k, m}(i)$ can be represented as sum of $A(c,m)$ events.
\begin{equation}
X_{k, m}(i) = \sum_{c\in \mathcal{N}_{k}(i)}{\mathbb{I}\left[A(c,m)\right]},
\end{equation}
leading to:
\begin{equation}
\mathbb{E}\left[X_{k, m}(i)\right] = \sum_{c\in \mathcal{N}_{k}(i)}{\mathbb{P}\left(A(c,m)\right)} 
\end{equation}
The probability in the right hand side is independent on $c$, since all edges are independent (up to the approximation of ignoring overlapping sets, which is a good approximation for large enough graphs), leading to:
\begin{equation}
\mathbb{E}\left[X_{k, m}(i)\right] = \begin{pmatrix}
n - 1 \\ k - 1
\end{pmatrix} \mathbb{P}\left(A(c,m)\right)
\end{equation}
The probability that a specific motif appears in a specific set of vertices  is just the sum of probabilities over all the isomorphs of $m$, that the specific isomorph appears. This probability is equal for all isomorphs, since all edges are independent, leading to:
\begin{equation}
\mathbb{E}\left[X_{k, m}(i)\right] = \begin{pmatrix}
n - 1 \\ k - 1
\end{pmatrix}N_{Iso}(m) \cdot p^{n_{e}(m)} \cdot (1-p)^{n_{max}(k) - n_{e}(m)}
\end{equation}

The estimate of  $N_{Iso}(m)$ is immediate and can be obtained by directly counting permutations. 

To test the fit between Eq. 4 and VDMC, we built $G(n, p)$ for multiple $n$ and used $p=\frac{1}{10}$, then compared our motif calculations to the theory. 
Figure 5 shows the log of the expected (internal bar) and observed (external bar) motif frequencies for graphs with 1,000 vertices. As one can see, the expected and observed values are equal (Chi square test non significant at the p=0.05 level for all motifs).

We have further performed extensive validations on both random graphs and small toy-graphs where the frequency of each motif can be computed analytically (e.g. cliques, regular Directed Acyclic Graphs (DAG), etc. ) and VDMC is always accurate. All examples are detailed in the GIT.
\begin{figure}
\includegraphics[width=0.75\textwidth]{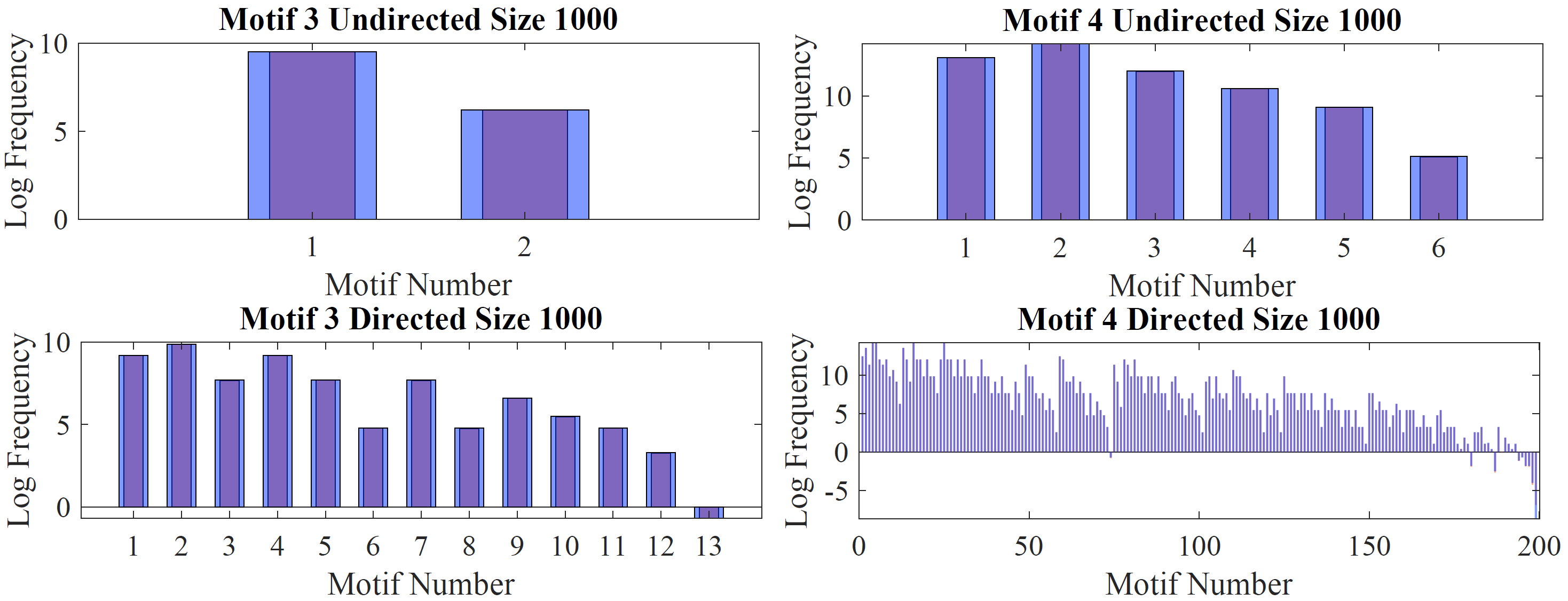}
\caption{Comparison between theory and VDMC results for motif 3 and 4 in directed and undirected graphs. Internal bars are theory and external bars are VDMC results. Upper plots are for undirected 3 and 4 motif respectively, and the lower plots are the same for directed motifs. }
\end{figure}

\section{Algorithm efficacy - simulated graphs}

\begin{figure}
\includegraphics[width=0.9\textwidth]{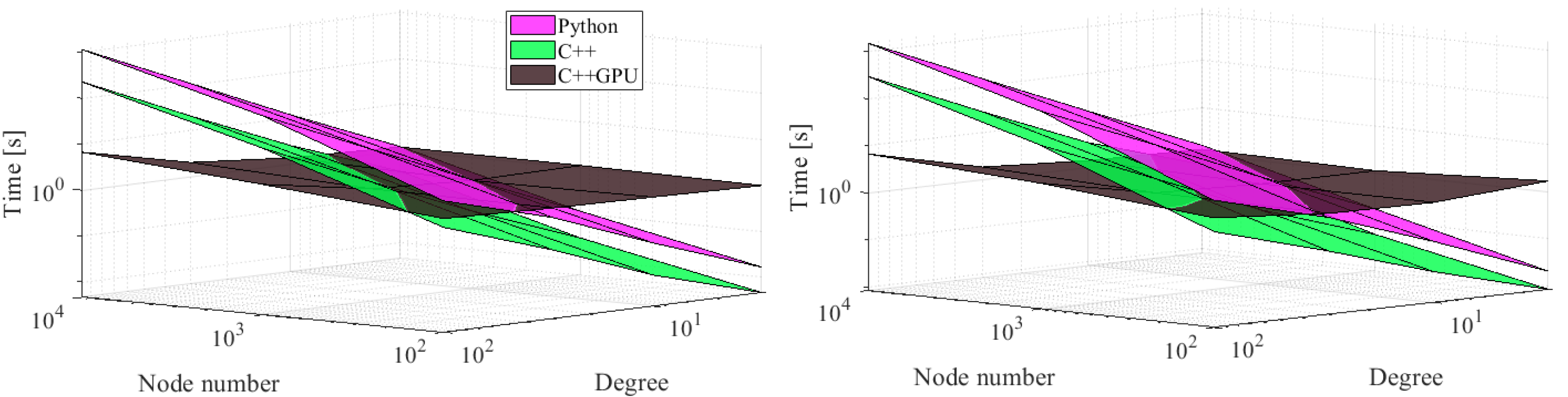}
\caption{Run time comparison for different implementation. The simulations were performed on Erd\H{o}s R\'enyi ($G(n, p)$) graphs, and the x and y axes are the vertex and edge numbers. The left plot is for undirected 4 motifs and the right plot is for directed 4 motifs.}
\label{fig:run times}
\end{figure}

\begin{figure}
\includegraphics[width=0.75\textwidth]{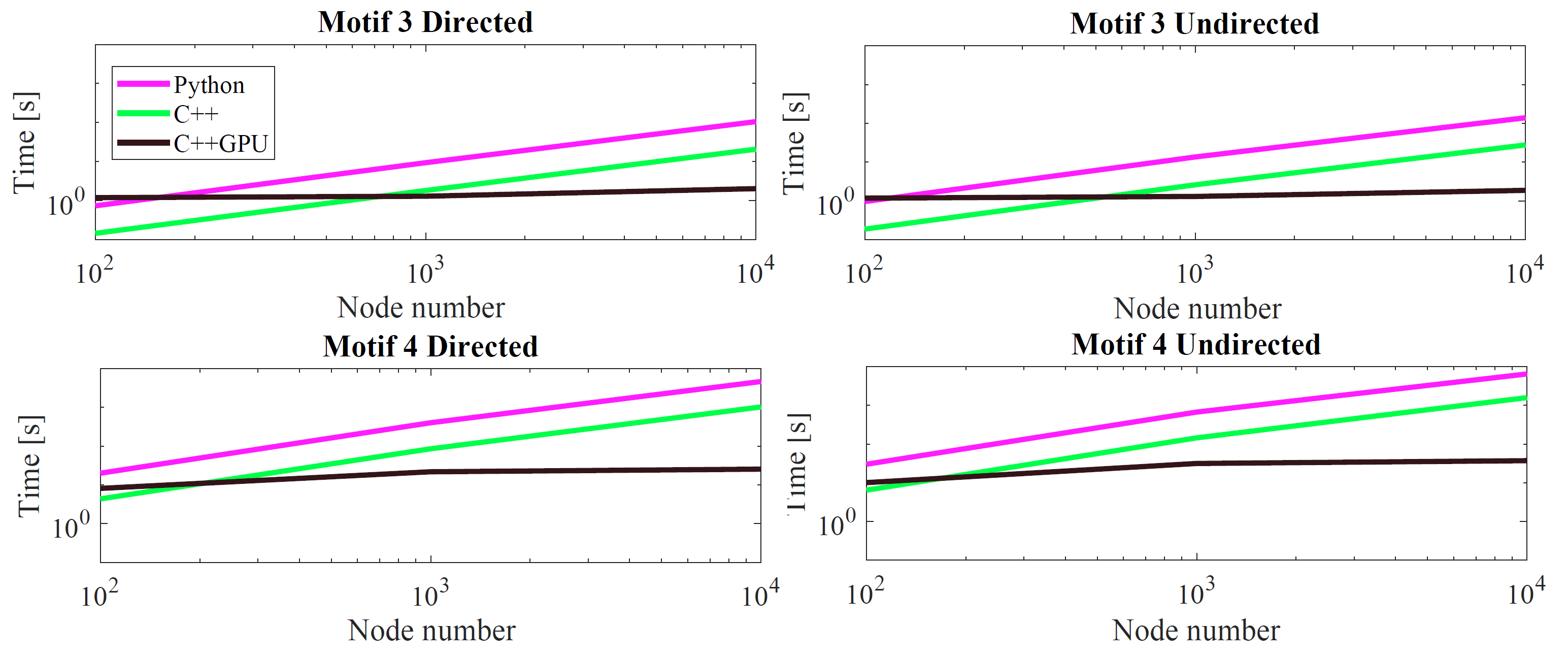}
\caption{Run time comparison for different implementation, for fixed average degree of 10.}
\label{fig:run times fixed degree}
\end{figure}

The cost of the proposed algorithm is $O(|V|*<k^3>)$ for $4-motifs$ and $O(|V|*<k^2>)$ for $3-motifs$, which is proportional to  the total number of motifs, where $<k^3>$ is the average number of third neighbors. For an Erd\H{o}s R\'enyi graph, this is closely equivalent to $O(|E|^3/|V|^2)$. However, for fat-tailed distributions, this can be much higher. Note that the efficient implementation ensures that the constants in this analysis are not large. In practice, the computational cost is determined by the implementation. The computational cost of the same algorithm in C++ is approximately 10 times more efficient than its parallel in Python (Fig \ref{fig:run times} and \ref{fig:run times fixed degree}). 

Moreover, following the massive parallelization, the cost of GPU based application is not sensitive to the number of vertices, and not to the average degree, as long as not all threads are filled. Thus, up to thousands of vertices and hundreds of thousands of edges, the cost of GPU based sub-graph enumeration can be treated as constant. Even for larger real-world graphs, the cost of the GPU application is of reasonable cost, as further detailed.
Note that there is an initial cost to accessing GPU. Thus, for small graphs, the computational cost of GPU based applications is in practice higher than CPU based applications. 

\section{Algorithm efficacy - real world networks}

Simulated graphs typically  differ from real world networks. Real world networks often have scale free degree distribution, and as such may be computationally expensive. VDMC can handle very high degrees through the division of the $k-BFS$ for high degree vertices into parallel computations, as mentioned in the GPU implementation. 
Several real datasets were used to test our algorithm and to compare it to previous algorithms. These datasets and their properties are summarized in table number \ref{table:Real Datasets}. The current state of the art for local subgraph counting is through
local homomorphism counting, and process local homomorphism counting by natural joins with group-by and aggregation in a distributed system DISC on top of Spark (see \cite{zhang} for more details). The elapsed time for each algorithm for 3 and $4-motifs$ may be found in table number \ref{table:Comparison VDMC with DISC}. VDMC is typically 5-10 time slower than DISC, but  VDMC is run on a single Tesla V100-SXM2-32GB card, while DISC is run on a 16 machine Spark system. Moreover, DISC only count undirected motifs, and VDMC counts directed motifs.

\begin{table}[h!]
\caption{Real world Datasets. The first column is the dataset name. The second column is the notation used in the text, followed by the number of vertices and edges, and whether the dataset is directed or not}
\label{table:Real Datasets}
\begin{center}
\begin{tabular}{ |c c|c|c|c| }
\hline
 \backslashbox{Datasets}{Properties}& Notation & |V| & |E|& Is Directed  \\
 \Xhline{2\arrayrulewidth}
 \multirow{2}{*}{web-BerkStan} & WBD & \multirow{2}{*}{$6.9*10^5$}  & $7.6*10^6$ & True   \\ 
 \cline{2-2}\cline{4-5}
 & WB && $6.6*10^6$ & False \\
 \hline
 as-Skitter & AS &  $1.7*10^6$ & $1.1*10^7$ & False  \\
 \hline
 \multirow{2}{*}{soc-LiveJournal} & LJD & \multirow{2}{*}{$4.8*10^6$} & $6.9*10^7$ & True  \\
 \cline{2-2}\cline{4-5}
 & LJ && $4.3*10^7$ & False \\
  \hline
 com-Orkut & OK &  $3.1*10^6$ & $1.2*10^8$ & False  \\
  \hline
\end{tabular}
\end{center}
\end{table}

\begin{table}[h!]
\caption{Comparison between VDMC and DISC running times in seconds for 3 and $4-motifs$. Note that VDMC is run on a single card, while DISC is run on a 16 machine Spark system. Moreover, DISC only counts undirected motifs, and VDMC counts directed motifs. }
\label{table:Comparison VDMC with DISC}
\begin{center}
\hspace*{-8pt}\makebox[\linewidth][c]{
\begin{tabular}{ |c|c| c| c| c|}
\hline
 \multirow{2}{*}{\backslashbox{Datasets}{Elapse Time (seconds)}} & \multirow{2}{*}{Notation} &\multicolumn{2}{c|}{VDMC} & DISC \\
 \cline{3-5}
 & & $3-Motif$ & $4-Motif$ & $4-Motif$ \\
 \Xhline{2\arrayrulewidth}
 \multirow{2}{*}{web-BerkStan} & WBD & 68 & 23736 & ---    \\ 
 \cline{2-5}
 & WB & 76 & 30315 & 659\\
 \hline
 as-Skitter & AS & 154 &6968  & 713  \\
 \hline
 \multirow{2}{*}{soc-LiveJournal} & LJD & 635 & 10882 & ---   \\
 \cline{2-5}
 & LJ & 574 & 4645 & 953 \\
  \hline
 com-Orkut & OK & 1628 & 28730 & 5043    \\
 \hline
\end{tabular}
}
\end{center}
\end{table}

\section{Other tools available  in the current toolbox}
The CSR format allows for efficient computation of multiple features, beyond the motif counting discussed here, we developed multiple other simpler measures using the same formalism, including:
\begin{itemize}
    \item K-Cores - the maximal sub-graph, where each vertex has a degree of at least $k$ in the sub-graph (\cite{dorogovtsev2006k}).
    \item The normalized distance distribution for each vertex (i.e. the fraction of vertices with a distance of $1,2....$ from a given vertex.
    \item Attraction Bassin hierarchy (\cite{muchnik2007self}).
    Average neighbor degree
    \item Page Rank (\cite{page1999pagerank}).
    \item Flow - a hierarchy measure that approximates topological sorting for graphs with cycles(\cite{rosen2014directionality}).
\end{itemize}
All these measures are available in the Github with the motif counting algorithm.

\section{Discussion}
We have presented a highly efficient distributed algorithm to count vertex participation in 3 and 4 motifs, and shown the accuracy and efficacy of the algorithm. The memory cost is simply the number of edges, and the CPU cost is precisely the number of motifs counted since each motif is counted once and only once. This algorithm contains an efficient cache aware data structure. The same could be extended to counting motifs for edges, rather than vertices. This change is minimal and only requires updating edges and not vertices once a motif was counted. The proposed algorithm can also be easily distributed among different GPUs/CPUs, by simply sending chunks of vertices in the root of the BFS to different GPUs/CPUs. Finally, while we have proposed an algorithm for 3 and 4 motifs, the current claims and data structure are appropriate for 5 motifs too.

This algorithm has many possible applications, including among many others, sub-graph isomorphism algorithms, topology-based graph machine learning, as well as descriptive tools for network analysis. As the networks studied grow exponentially, such tools are getting critical to enlarge existing network topology methods to larger graphs. 

The main caveat of the current method is its limitation to a given motif size, in contrast with methods that have been developed to detect any specific sub-graph ( \cite{kashtan2004efficient,kashani2009kavosh,wernicke2006fanmod,melckenbeeck2018efficiently,song2015method,koskas2011nemo}). However, the enumeration of all sub-graph a given size requires only one pass on each set of vertices that can compose such a motif. As such, the current code is typically as efficient for all motifs as for the task of finding a specific sub-graph of a given size.
\newpage
\bibliographystyle{ACM-Reference-Format}
\bibliography{sample-base}
\section{Appendix I - GPU Changes}

To further accelerate the motif counting code, a GPU based version  was
developed using the CUDA library. In the parallel version, the frequencies of all motifs for each $k-BFS(i)$ are computed in parallel. This requires running each of those calculations independently of the others. Given the lemmas proved above, each $k-BFS(i)$ can be computed separately. 

The following modifications were required for the GPU implementation:
\begin{enumerate}
\item \textbf{GPU functions }are written outside the main bodies of code
(MotifCalculator,MotifUtils, CacheGraph), as they cannot be member
functions. They are instead re-written as stand-alone functions within
the file.
\item \textbf{Global variables} are used to communicate between the class
methods, which provide the CPU-side pre- and post-processing of the
results, and the GPU functions.
\item \textbf{Atomic add} is used to update the motif counters, so that
the GPU threads won't interfere with one another.
\item \textbf{All data} which is used in the GPU must be copied into GPU
memory. Additionally, the data is prefetched asynchronously to improve
the code's performance.
\end{enumerate}

To run the GPU code, some requirements must be met, in both
software and hardware.
\begin{itemize}
\item The GPU must be an NVIDIA GPU of computing capability 3.5 or higher.
\item Version 8.0 or higher of the CUDA Toolkit must be installed and in the PATH
of the system. 
\item The GPU drivers must be compatible with the CUDA Toolkit.
\end{itemize}
Specifically, the code was tested on a system with the following specs:
\begin{itemize}
\item Ubuntu 20.04.2 LTS
Linux distribution
\item A Tesla V100-SXM2-32GB
with compute capability 7.0
\item Each of the following CUDA Toolkit versions: 8.0, 9.2, 10.1, 10.2
\item GPU driver versions 396.26 and 440.33.01
\end{itemize}
\end{document}